\documentclass[prd,superscriptaddress,showpacs,nofootinbib,amsmath,amssymb]{revtex4}
\usepackage{slashed,graphicx,color,axodraw,hyperref}
\begin{document}
\title{The Oblique Corrections from Heavy Scalars in Irreducible Representations}
\author{Hong-Hao Zhang}
\affiliation{%
Department of Physics, Tsinghua University, Beijing 100084, China}

\author{Wen-Bin Yan}
\affiliation{%
Department of Physics, Tsinghua University, Beijing 100084, China}

\author{Xue-Song Li}
\affiliation{%
Science College, Hunan Agricultural University, Changsha 410128,
China}

\begin{abstract}
The contributions to $S$, $T$, and $U$ from heavy scalars in any
irreducible representation of the electroweak gauge group
$SU(2)_L\times U(1)_Y$ are obtained. We find that in the case of a
heavy scalar doublet there is a slight difference between the $S$
parameter we have obtained and that in previous works.
\end{abstract}
\pacs{12.15.Lk, 12.38.Bx} \maketitle
%%%%%%%%%%%%%%%%%%%%%%%%%%%%%%%%%%%%%%%%%%%%%%%%%%%%%%
\section{Introduction}
In spite of its tremendous success, the standard model (SM) has
several drawbacks. The Higgs particle has not yet been found in
experiments; on the other hand, the SM suffers unnaturalness and
triviality from a theoretical point of view. Thus, the SM may not be
correct, or at least it is just an effective theory at the
electroweak scale. There are many new physics possibilities beyond
the SM. Although we do not know whether nature really behaves like
one of them or not, we can estimate their effects on the current
electroweak precision measurements. Peskin and Takeuchi's
$S$-$T$-$U$-formalism is a practical way to do this job
\cite{Peskin:1990zt}. Since the current SM parameter fits indicate
that $S$ and $T$ are small negative numbers, and $U$ is also close
to zero \cite{Yao:2006px}, those new physics models which give large
positive contributions to $S$ and $T$ are presumably excluded. Thus,
the oblique correction parameters $S$, $T$, and $U$ are often used
to judge whether a new model is compatible with experiments or not.
If the SM is not a full theory, there will be new heavy particles
above the electroweak scale. Provided the new particles feel the
electroweak interactions, they should give corrections to $S$, $T$,
and $U$ no matter whether they are fermions, scalars, or gauge
bosons. The contributions of heavy fermions and scalars to the
oblique correction parameters has been studied extensively in the
literature
\cite{Peskin:1990zt,Georgi:1991ci,Dugan:1991ck,Gates:1991uu,Luty:1992fe,Li:1992dt,
Bhattacharyya:1993zy,Inami:1994nj,Bhattacharya:1994bj,Popovic:2000dx,Peskin:2001rw,
He:2001tp,Wells:2005vk}. And in a previous work we have obtained the
oblique corrections from fermions in higher-dimensional
representations of the electroweak gauge group $SU(2)_L\times
U(1)_Y$ \cite{Zhang:2006de}.

In this paper, we obtain the oblique corrections from scalars in any
irreducible representation of $SU(2)_L\times U(1)_Y$. We begin by
computing the contributions to $S$, $T$ and $U$ from a heavy scalar
doublet which has no vacuum expectation value (VEV). The $T$
parameter from this scalar doublet is exactly the same as the
left-handed fermion doublet contribution, which has been shown first
in Ref.\cite{Li:1992dt}. However, there is a slight difference
between the $S$ parameter we have obtained and that in
Ref.\cite{Li:1992dt}. This difference may come from the convention
inconsistency of Ref.\cite{Li:1992dt}. We then proceed to obtain the
contributions to $S$, $T$ and $U$ from higher-dimensional scalar
representations. Throughout this paper, we assume that the heavy
scalar multiplet we consider has never acquired a VEV, and there is
a slight mass non-degeneracy in the scalar multiplet which may be
caused by unknown ultraviolet completions.

This paper is organized as follows. In Sec. \ref{section-doublet} we
compute the contributions to $S$, $T$ and $U$ from a heavy complex
scalar doublet. In Sec. \ref{higher-dim} we obtain the explicit
expressions of the oblique corrections from a $(2j+1)$-dimensional
scalar multiplet. Conclusions and discussions are given in Sec.
\ref{conclusions}. In Appendices \ref{appendix-def} and
\ref{general-vacpol}, some basic definitions and relations are
listed, and general scalar vacuum polarization amplitudes that help
to calculate vector-boson self energies are given.

\section{A Heavy Scalar Doublet\label{section-doublet}}
To begin with, let us consider a heavy complex scalar doublet
$$\Phi=\begin{pmatrix}\phi_1\\\phi_2\end{pmatrix}$$
which belongs to the representation $(2,~Y)$ of $SU(2)_L\times
U(1)_Y$. The heavy scalar fields $\phi_1$ and $\phi_2$ can acquire
their masses from either bare mass terms or some other fields' VEVs
much larger than the electroweak scale. In general, the doublet
$\Phi$ is non-degenerate, and the masses of $\phi_1$ and $\phi_2$
are denoted by, respectively, $m_1$ and $m_2$. These scalars couple
to the electroweak gauge bosons via
\begin{eqnarray}
\mathcal{L}_{int}&=&i\Phi^\dag\bigg[\frac{e}{\sqrt{2}s}(W_\mu^+T^++W_\mu^-T^-)
+\frac{e}{sc}Z_\mu(T^3-s^2Q)+eA_\mu Q\bigg](\partial^\mu\Phi)
-i(\partial^\mu\Phi^\dag)\bigg[\frac{e}{\sqrt{2}s}(W_\mu^+T^++W_\mu^-T^-)\nonumber\\
&& +\frac{e}{sc}Z_\mu(T^3-s^2Q)+eA_\mu Q\bigg]\Phi
+\Phi^\dag\bigg[\frac{e}{\sqrt{2}s}(W_\mu^+T^++W_\mu^-T^-)
+\frac{e}{sc}Z_\mu(T^3-s^2Q)+eA_\mu
Q\bigg]^2\Phi\;,\label{interaction-doublet}
\end{eqnarray}
where $c\equiv\cos\theta_W$, $s\equiv\sin\theta_W$,
$W_\mu^\pm\equiv\frac{1}{\sqrt{2}}(W_\mu^1\mp iW_\mu^2)$,
$T^\pm\equiv T^1\pm iT^2$, $Q\equiv T^3+Y$, and $T^a\equiv\tau^a/2$
with $\tau^a$ ($a=1,2,3$) being Pauli matrices. In order to
calculate the oblique corrections from this heavy scalar doublet, it
is sufficient to calculate its contributions to the self energies of
$W$ and $Z$ bosons. The relevant Feynman rules can be easily read
off from Eq.\eqref{interaction-doublet} as follows:
\begin{itemize}
  \item $V\phi\phi$ vertices
\begin{eqnarray*}
W_\mu^+(-p-q)\phi_2(p)\phi_1^\dag(q)&:&~i\frac{e}{\sqrt{2}s}(p-q)^\mu\\
W_\mu^-(-p-q)\phi_1(p)\phi_2^\dag(q)&:&~i\frac{e}{\sqrt{2}s}(p-q)^\mu\\
Z_\mu(-p-q)\phi_1(p)\phi_1^\dag(q)&:&~i\frac{e}{sc}\frac{[1-(1+2Y)s^2]}{2}(p-q)^\mu\\
Z_\mu(-p-q)\phi_2(p)\phi_2^\dag(q)&:&~i\frac{e}{sc}\frac{[-1+(1-2Y)s^2]}{2}(p-q)^\mu
\end{eqnarray*}
with all momenta incoming.
  \item $VV\phi\phi$ vertices
\begin{eqnarray*}
W_\mu^+W_\nu^-\phi_1\phi_1^\dag&:&~i\frac{e^2}{2s^2}g^{\mu\nu}\\
W_\mu^+W_\nu^-\phi_2\phi_2^\dag&:&~i\frac{e^2}{2s^2}g^{\mu\nu}\\
Z_\mu
Z_\nu\phi_1\phi_1^\dag&:&~i\frac{e^2}{s^2c^2}\frac{[1-(1+2Y)s^2]^2}{2}g^{\mu\nu}\\
Z_\mu
Z_\nu\phi_2\phi_2^\dag&:&~i\frac{e^2}{s^2c^2}\frac{[1-(1-2Y)s^2]^2}{2}g^{\mu\nu}
\end{eqnarray*}
\end{itemize}
Using these Feynman rules, the contributions of the scalar doublet
to the $W$-boson self energy can be written in terms of the building
blocks $i\,\Pi_\Omega^{\mu\nu}(q^2,m^2)$ and
$i\,\Pi_\Phi^{\mu\nu}(q^2,m_1^2,m_2^2)$ defined in Appendix
\ref{general-vacpol} as follows
\begin{eqnarray}
i\,\Pi_{WW}^{\mu\nu}(q^2)&=&\frac{e^2}{2s^2}i\,\Pi_\Omega^{\mu\nu}(q^2,m_1^2)
+\frac{e^2}{2s^2}i\,\Pi_\Omega^{\mu\nu}(q^2,m_2^2)
+(\frac{e}{\sqrt{2}s})^2i\,\Pi_\Phi^{\mu\nu}(q^2,m_1^2,m_2^2)\nonumber\\
&=&i\frac{e^2}{2s^2}\frac{g^{\mu\nu}}{(4\pi)^2}\bigg[m_1^2\log\frac{m_1^2}{\mu^2}
+m_2^2\log\frac{m_2^2}{\mu^2}-2f_2(m_1^2,m_2^2)
+q^2[-\frac{1}{3}\Upsilon+2f_1(m_1^2,m_2^2)]\bigg]\nonumber\\
&&+(q^\mu q^\nu\mbox{~terms})+\mathcal{O}(q^4)\;,\label{Pi-WW}\\[2ex]
i\,\Pi_{ZZ}^{\mu\nu}(q^2)&=&\frac{e^2}{s^2c^2}\frac{[1-(1+2Y)s^2]^2}{2}
i\,\Pi_\Omega^{\mu\nu}(q^2,m_1^2)
+\frac{e^2}{s^2c^2}\frac{[1-(1-2Y)s^2]^2}{2}i\,\Pi_\Omega^{\mu\nu}(q^2,m_2^2)\nonumber\\
&&+\bigg(\frac{e}{sc}\frac{[1-(1+2Y)s^2]}{2}\bigg)^2i\,\Pi_\Phi^{\mu\nu}(q^2,m_1^2,m_1^2)
+\bigg(\frac{e}{sc}\frac{[-1+(1-2Y)s^2]}{2}\bigg)^2
i\,\Pi_\Phi^{\mu\nu}(q^2,m_2^2,m_2^2)\nonumber\\
&=&i\frac{e^2}{s^2c^2}\frac{g^{\mu\nu}}{(4\pi)^2}\frac{q^2}{12}\bigg[
(-2\Upsilon+\log\frac{m_1^2}{\mu^2}+\log\frac{m_2^2}{\mu^2})
-2s^2[-2\Upsilon+(1+2Y)\log\frac{m_1^2}{\mu^2}+(1-2Y)\log\frac{m_2^2}{\mu^2}]\nonumber\\
&&+s^4[-2(1+4Y^2)\Upsilon+(1+2Y)^2\log\frac{m_1^2}{\mu^2}+(1-2Y)^2\log\frac{m_2^2}{\mu^2}]
\bigg]+(q^\mu q^\nu\mbox{~terms})+\mathcal{O}(q^4)\;,\label{Pi-ZZ}
\end{eqnarray}
where $\mu$, $\Upsilon$, $f_1$ and $f_2$ have been defined in
Appendix \ref{general-vacpol}. Now, combining Eqs.\eqref{Pi-WW} and
\eqref{Pi-ZZ} with Eqs.\eqref{def-vacuum-pol},
\eqref{def-vacuum-pol2} and \eqref{def-vacuum-pol3} in Appendix
\ref{appendix-def}, we have
\begin{eqnarray}
&&\Pi_{11}(0)=\frac{1}{2(4\pi)^2}\bigg[m_1^2\log\frac{m_1^2}{\mu^2}
+m_2^2\log\frac{m_2^2}{\mu^2}-2f_2(m_1^2,m_2^2)\bigg]\;,\nonumber\\
&&\Pi_{11}^\prime(0)=\frac{1}{2(4\pi)^2}
\bigg[-\frac{1}{3}\Upsilon+2f_1(m_1^2,m_2^2)\bigg]\;,\nonumber\\
&&\Pi_{33}(0)=0\;,\qquad \Pi_{3Q}(0)=0\;,\label{Pi-1-3-Q}\\
&&\Pi_{33}^\prime(0)=\frac{1}{12(4\pi)^2}\bigg[-2\Upsilon+\log\frac{m_1^2}{\mu^2}
+\log\frac{m_2^2}{\mu^2}\bigg]\;,\nonumber\\
&&\Pi_{3Q}^\prime(0)=\frac{1}{12(4\pi)^2}
\bigg[-2\Upsilon++(1+2Y)\log\frac{m_1^2}{\mu^2}+(1-2Y)\log\frac{m_2^2}{\mu^2}\bigg]
\;.\nonumber
\end{eqnarray}
Thus, from Eqs.\eqref{Pi-1-3-Q} and \eqref{def-STU}, we obtain
\begin{eqnarray}
&&S=-\frac{Y}{6\pi}\log\frac{m_1^2}{m_2^2}\;,\label{S-doublet}\\
&&T=\frac{1}{8\pi s^2c^2m_Z^2}\bigg[m_1^2\log\frac{m_1^2}{\mu^2}
+m_2^2\log\frac{m_2^2}{\mu^2}-2f_2(m_1^2,m_2^2)\bigg]\;,\label{T-doublet}\\
&&U=\frac{1}{\pi}\bigg[f_1(m_1^2,m_2^2)-\frac{1}{12}(\log\frac{m_1^2}{\mu^2}
+\log\frac{m_2^2}{\mu^2})\bigg]\;,\label{U-doublet}
\end{eqnarray}
from which, the explicit expressions for $T$ and $U$ can be obtained
by figuring out the functions $f_1$ and $f_2$ as follows:
\begin{eqnarray}
&&T=\frac{1}{16\pi s^2c^2m_Z^2}
\bigg(m_1^2+m_2^2-\frac{2m_1^2m_2^2}{m_1^2-m_2^2}\log\frac{m_1^2}{m_2^2}\bigg)\;,
\label{T-doublet-2}\\
&&U=\frac{1}{12\pi}\bigg[-\frac{5m_1^4-22m_1^2m_2^2+5m_2^4}{3(m_1^2-m_2^2)^2}
+\frac{m_1^6-3m_1^4m_2^2-3m_1^2m_2^4+m_2^6}{(m_1^2-m_2^2)^3}
\log\frac{m_1^2}{m_2^2}\bigg]\;.
\end{eqnarray}
Comparing them with the results of Ref.\cite{Peskin:1990zt}, it is
interesting to note that the contribution to the $T$ parameter from
the heavy scalar doublet is exactly the same as that from one heavy
SM-like fermion doublet. And the contribution of the scalar doublet
to the $U$ parameter differs only by a factor of $1/2$ from that of
one SM-like fermion doublet. Moreover, except for an additional
positive piece $1/(6\pi)$ in the SM-like fermion case, the
difference of the scalar and SM-like fermion doublets' contributions
to $S$ is also a factor of $1/2$ \footnote{Note that in this paper
the hypercharge $Y$ is defined by $Q=T_3+Y$, while in
Ref.\cite{Peskin:1990zt} $Q=T_3+\frac{Y}{2}$. Thus, if taken this
paper's notation, the $Y$ appearing in the original expressions in
the Ref.\cite{Peskin:1990zt} must be replaced by $2Y$.}.

In Ref.\cite{Li:1992dt}, the contributions to $S$ and $T$ from a
heavy scalar doublet with hypercharge $Y=-1/2$ are given. The
expression for the $T$ parameter in our paper, {\it i.e.}
Eq.\eqref{T-doublet-2}, coincides with that in Ref.\cite{Li:1992dt}.
However, for the case of $Y=-1/2$ in Eq.\eqref{S-doublet}, we have
\begin{eqnarray}
S=\frac{1}{12\pi}\log\frac{m_1^2}{m_2^2}\;,\label{S-doublet-Y-minus1over2}
\end{eqnarray}
which differs by a factor of 2 from the result quoted in Eq.(18) of
Ref.\cite{Li:1992dt}. We reexamine Ref.\cite{Li:1992dt} and find
that this difference stems from the inconsistency of the convention
of Ref.\cite{Li:1992dt}. Obviously, Eq.(5) of Ref.\cite{Li:1992dt}
describes one of the self energy diagrams for the charged
$W_\mu^\pm$ while Eq.(6) of that paper corresponds to $W_\mu^a$
($a=1,2,3$). If one adds up these two diagrams, then the convention
is inconsistent. Were Eq.(18) of Ref.\cite{Li:1992dt} imprecise,
some subsequent literature directly quoting the result should be
modified as well. For instance, Eq.(8) of
Ref.\cite{Bhattacharyya:1993zy} has also an over-multiplying factor
of $1/2$.

\section{Higher-Dimensional Representations\label{higher-dim}}
In this section we proceed to give the contributions to $S$, $T$ and
$U$ from scalars in more general representations. Consider a heavy
complex scalar multiplet $\Phi$ with quantum numbers of
$SU(2)_L\times U(1)_Y$ as follows
\begin{eqnarray}
\Phi=\begin{pmatrix} \phi_j\\
\phi_{j-1}\\
\vdots\\
\phi_{-j}\end{pmatrix}\sim(2j+1,~Y)\;.
\end{eqnarray}
The mass of $\phi_l$ is denoted by $m_l$ for $l$ running from $-j$
to $j$. The interactions between these scalars and the electroweak
gauge bosons is of the same form as Eq.\eqref{interaction-doublet},
but at this point
\begin{eqnarray}
&&(T^{\pm})_{kl}=\sqrt{(j\mp l)(j\pm l+1)}\delta_{k\mp1,l}\;,\nonumber\\
&&(T^3)_{kl}=l\delta_{kl}\;,\qquad Q_{kl}=(Y+l)\delta_{kl}\;,\qquad
-j\leqslant k,l\leqslant j\;.
\end{eqnarray}
Likewise, we compute their contributions to $S$, $T$, and $U$
resulting in
\begin{eqnarray}
&&S=-\frac{Y}{3\pi}\sum_{l=-j}^j\,l\,\log\frac{m_l^2}{\mu^2}\;,\label{S-N-plet}\\
&&T=\frac{1}{4\pi
s^2c^2m_Z^2}\bigg[\sum_{l=-j}^j(j^2+j-l^2)m_l^2\,\log\frac{m_l^2}{\mu^2}
-\sum_{l=-j}^{j-1}(j-l)(j+l+1)f_2(m_{l}^2,m_{l+1}^2)\bigg]\;,\label{T-N-plet}\\
&&U=\frac{1}{\pi}\bigg[\sum_{l=-j}^{j-1}(j-l)(j+l+1)f_1(m_{l}^2,m_{l+1}^2)
-\sum_{l=-j}^j\frac{l^2}{3}\log\frac{m_l^2}{\mu^2}\bigg]\;.\label{U-N-plet}
\end{eqnarray}
It is easily checked that, by taking $j=1/2$, Eqs.\eqref{S-N-plet},
\eqref{T-N-plet} and \eqref{U-N-plet} agree with
Eqs.\eqref{S-doublet}, \eqref{T-doublet} and \eqref{U-doublet}.
Comparing the above results with those of Ref.\cite{Zhang:2006de},
we can see that the equality of the contributions to the $T$
parameter from a scalar $(2j+1)$-plet and a SM-like fermion
$(2j+1)$-plet is just a coincidence for $j=1/2$. It is interesting
to find that the contribution of a scalar $(2j+1)$-plet to the $S$
parameter is always one half of that of a vectorlike fermion
$(2j+1)$-plet, and the contribution of a scalar $(2j+1)$-plet to the
$U$ parameter is always one half of that of a SM-like fermion
$(2j+1)$-plet. Note that in the degenerate mass limit these
expressions for $S$, $T$ and $U$ all vanish, and thus the current
precision electroweak measurements cannot rule out a heavy
degenerate scalar multiplet. Moreover, the small experimental values
of $S$, $T$ and $U$ could impose rather stringent constraints on the
mass non-degeneracy of the scalar multiplet.

\section{Conclusions\label{conclusions}}

In this paper, we have obtained the oblique corrections from heavy
scalars in any irreducible representations of the electroweak gauge
group $SU(2)_L\times U(1)_Y$. Our expression for the $S$ parameter
in the case of a heavy scalar doublet with $Y=-1/2$ is slightly
different from that in Ref.\cite{Li:1992dt}. We have pointed out
that the convention inconsistency of that paper most likely causes
the disagreement. We have used dimensional regularization to perform
scalar loop calculations with an implicit assumption that possible
quadratic divergent terms are exactly canceled in the expressions of
$S$, $T$ and $U$, which must be true for the purpose of the oblique
correction formalism. The scalar multiplet we considered above has
no VEV and no mixing with the SM Higgs doublet. We have shown that
such a heavy degenerate scalar multiplet is not excluded by the
current $S$-$T$-$U$ fits.

\appendix
\section{Basic Definitions and Relations\label{appendix-def}}
In this appendix we briefly recapitulate some basic definitions and
relations given in Ref.\cite{Peskin:1990zt}. The vacuum-polarization
amplitudes are defined as follow:
\begin{eqnarray}
i\Pi_{XY}^{\mu\nu}(q^2)&=&ig^{\mu\nu}\Pi_{XY}(q^2)+(q^\mu q^\nu
\mbox{terms}) \nonumber\\
&\equiv&\int d^4xe^{-iqx}\langle
J_X^\mu(x)J_Y^\nu(0)\rangle\;,\label{def-vacuum-pol}
\end{eqnarray}
where $(XY)=(11), (22), (33), (3Q),\mbox{and} (QQ)$. And
$\Pi_{XY}^\prime(q^2)$ is defined by
\begin{eqnarray}
\Pi_{XY}(q^2)\equiv
\Pi_{XY}(0)+q^2\Pi_{XY}^\prime(q^2)\;.\label{def-vacuum-pol2}
\end{eqnarray}
The relations between the one-particle irreducible (1PI)
self-energies of the gauge bosons and the vacuum-polarization
amplitudes are given by
\begin{eqnarray}
&&\Pi_{AA}=e^2\Pi_{QQ},\quad
\Pi_{ZA}=\frac{e^2}{sc}(\Pi_{3Q}-s^2\Pi_{QQ}),\nonumber\\
&&\Pi_{ZZ}=\frac{e^2}{s^2c^2}(\Pi_{33}-2s^2\Pi_{3Q}+s^4\Pi_{QQ}),\label{def-vacuum-pol3}\\
&&\Pi_{WW}=\frac{e^2}{s^2}\Pi_{11}\;,\nonumber
\end{eqnarray}
where $c\equiv\cos\theta_W$, $s\equiv\sin\theta_W$, and e is the
coupling constant of the electromagnetic interaction. The three
oblique correction parameters $S$, $T$ and $U$ are defined
respectively by
\begin{eqnarray}
&&\alpha S\equiv 4e^2[\Pi_{33}^\prime(0)-\Pi_{3Q}^\prime(0)]\;,\nonumber\\
&&\alpha
T\equiv\frac{e^2}{s^2c^2m_Z^2}[\Pi_{11}(0)-\Pi_{33}(0)]\;,\label{def-STU}\\
&&\alpha U\equiv
4e^2[\Pi_{11}^\prime(0)-\Pi_{33}^\prime(0)]\;,\nonumber
\end{eqnarray}
where $\alpha\equiv e^2/(4\pi)$ is the fine-structure constant.

\section{Generic Complex Scalar Vacuum Polarization Amplitudes\label{general-vacpol}}
\begin{figure}[!h]
\begin{center}
\begin{picture}(200,100)(-100,0)
\SetColor{Red} \DashCArc(50,25)(15,0,360){3} \SetColor{Blue}
\Photon(5,25)(35,25){3}{4} \Photon(65,25)(95,25){3}{4}
\put(6,19){\vector(1,0){20}} \Text(13,17)[t]{$q$} \SetColor{Brown}
\Vertex(35,25){2} \Vertex(65,25){2} \Text(38,35)[rb]{$\phi_1^\dag$}
\Text(62,35)[lb]{$\phi_1$} \Text(38,15)[rt]{$\phi_2$}
\Text(62,15)[lt]{$\phi_2^\dag$} \Text(10,30)[rb]{$\mu$}
\Text(90,30)[lb]{$\nu$} \Text(-90,25)[l]{$\displaystyle
i\,\Pi_\Phi^{\mu\nu}(q^2,m_1^2,m_2^2)$}\Text(-5,25)[r]{$\equiv$}
%%%%%%%%%%%%%%%%%%%%%%%%%%%%%%%%%%%%%%%%%%%%%%%%%%%%%%%%%%%%%%%%%%
\SetColor{Red} \DashCArc(50,80)(15,0,360){3} \SetColor{Blue}
\Photon(10,65)(50,65){2}{5} \Photon(90,65)(50,65){2}{5}
\Text(33,85)[rt]{$\phi$} \Text(67,85)[lt]{$\phi^\dag$}
\put(12,60){\vector(1,0){20}} \Text(20,58)[t]{$q$} \SetColor{Brown}
\Vertex(50,65){2} \Text(15,70)[rb]{$\mu$} \Text(85,70)[lb]{$\nu$}
\Text(-90,75)[l]{$\displaystyle i\,\Pi_\Omega^{\mu\nu}(q^2,m^2)$}
\Text(-5,75)[r]{$\equiv$}
\end{picture}
\end{center}
\caption[1]{\label{scalar-vacuum-pol} Two types of generic complex
scalar vacuum polarization amplitudes,
$i\,\Pi_\Omega^{\mu\nu}(q^2,m^2)$ and
$i\,\Pi_\Phi^{\mu\nu}(q^2,m_1^2,m_2^2)$, which are related to
vector-boson self energies. The subscripts $\Omega$ and $\Phi$ are
chosen due to the shape of the respective Feynman diagram.}
\end{figure}
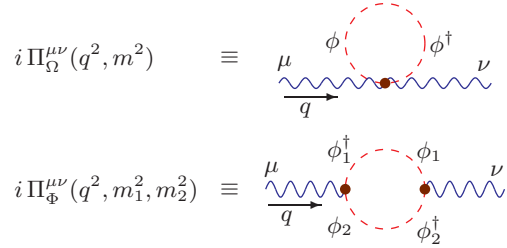
%%%%%%%%%%%%%%%%%%%%%%%%%%%%%%%%%%%%%%%%%%%%%%%%%%%%%%%%%%%%%%%%%%%%%%%%%%%
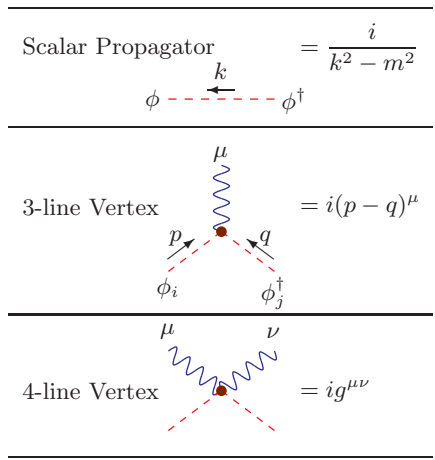
\begin{figure}[!h]
\begin{center}
\begin{picture}(150,170)(-5,0)
\put(-5,0){\line(1,0){160}} \SetColor{Blue}
\Photon(75,25)(55,40){3}{4} \Photon(95,40)(75,25){3}{4}
\SetColor{Red} \DashLine(75,25)(55,10){3} \DashLine(75,25)(95,10){3}
\SetColor{Brown} \Vertex(75,25){2} \Text(55,45)[b]{$\mu$}
\Text(95,45)[b]{$\nu$} \Text(0,25)[l]{4-line Vertex}
\Text(105,25)[l]{$=ig^{\mu\nu}$}
%%%%%%%%%%%%%%%%%%%%%%%%%%%%%%%%%%%%%%%%%%%%%%%%%%%%%
\put(-5,54){\line(1,0){160}} \SetColor{Blue}
\Photon(75,85)(75,110){3}{4} \SetColor{Red}
\DashLine(75,85)(55,70){3} \DashLine(75,85)(95,70){3}
\SetColor{Brown} \Vertex(75,85){2} \SetColor{Black}
\put(55,75){\vector(4,3){10}} \put(95,75){\vector(-4,3){10}}
\Text(58,80)[b]{$p$} \Text(92,80)[b]{$q$} \Text(55,68)[t]{$\phi_i$}
\Text(95,68)[t]{$\phi_j^\dag$} \Text(75,112)[b]{$\mu$}
\Text(0,95)[l]{3-line Vertex} \Text(105,95)[l]{$=i(p-q)^\mu$}
%%%%%%%%%%%%%%%%%%%%%%%%%%%%%%%%%%%%%%%%%%%%%%%%%%%%%%
\put(-5,125){\line(1,0){160}} \put(80,139){\vector(-1,0){10}}
\Text(75,143)[b]{$k$} \SetColor{Red} \DashLine(55,135)(95,135){3}
\Text(52,135)[r]{$\phi$} \Text(98,135)[l]{$\phi^\dag$}
\Text(0,155)[l]{Scalar Propagator} \Text(105,155)[l]{$\displaystyle
=\frac{i}{k^2-m^2}$} \put(-5,170){\line(1,0){160}}
\end{picture}
\end{center}
\caption[1]{\label{FeynRules} The defining Feynman Rules for the
generic complex scalar vacuum polarization amplitudes
$i\,\Pi_\Omega^{\mu\nu}(q^2,m^2)$ and
$i\,\Pi_\Phi^{\mu\nu}(q^2,m_1^2,m_2^2)$.}
\end{figure}
In this appendix we present some notations to simplify the
calculations of vector-boson self energies. Rather than computing
the specific complex scalar vacuum polarization amplitudes one by
one, it is convenient to compute, once and for all, the most general
ones which may give contributions to vector-boson self energies.
There are two types of generic complex scalar vacuum polarization
graphs as shown in Fig.\ref{scalar-vacuum-pol} with the Feynman
rules defined in Fig.\ref{FeynRules}. Then, performing dimensional
regularization gives
\begin{eqnarray}
i\,\Pi_\Omega^{\mu\nu}(q^2,m^2)&=&ig^{\mu\nu}\int\frac{d^4k}{(2\pi)^4}
\frac{i}{k^2-m^2}\rightarrow
ig^{\mu\nu}\mu^{4-d}\int\frac{d^dk}{(2\pi)^d}
\frac{i}{k^2-m^2}\nonumber\\
&=&-ig^{\mu\nu}\frac{m^2}{(4\pi)^2}(\Upsilon+1-\log\frac{m^2}{\mu^2})\;,
\label{iPi-Omega}\\
i\,\Pi_\Phi^{\mu\nu}(q^2,m_1^2,m_2^2)&=&\int\frac{d^4k}{(2\pi)^4}
\frac{(2k+q)^\mu(2k+q)^\nu}{(k^2-m_1^2)[(k+q)^2-m_2^2]}\rightarrow
\mu^{4-d}\int\frac{d^dk}{(2\pi)^d}
\frac{(2k+q)^\mu(2k+q)^\nu}{(k^2-m_1^2)[(k+q)^2-m_2^2]}\nonumber\\
&=&\frac{ig^{\mu\nu}}{(4\pi)^2}\bigg[ (m_1^2+m_2^2)(\Upsilon+1)
-2f_2(m_1^2,m_2^2)+q^2[-\frac{1}{3}\Upsilon+2f_1(m_1^2,m_2^2)]\bigg]\nonumber\\
&&+(q^\mu q^\nu\mbox{~terms})+\mathcal{O}(q^4)\;,\label{iPi-Phi}
\end{eqnarray}
where $\mu$ is an arbitrary mass scale parameter and the infinity
$\Upsilon\equiv2/(4-d)-\gamma+\log(4\pi)$, and where the functions
$f_1(m_1^2,m_2^2)$ and $f_2(m_1^2,m_2^2)$ are defined respectively
by
\begin{eqnarray}
&&f_1(m_1^2,m_2^2)\equiv\int_0^1dx\,x(1-x)\log\bigg[\frac{xm_1^2
+(1-x)m_2^2}{\mu^2}\bigg]\;,\label{f1}\\
&&f_2(m_1^2,m_2^2)\equiv\int_0^1dx[xm_1^2+(1-x)m_2^2]
\log\bigg[\frac{xm_1^2+(1-x)m_2^2}{\mu^2}\bigg]\;.\label{f2}
\end{eqnarray}
In particular, if $m_1=m_2=m$ in Eq.\eqref{iPi-Phi}, we have
\begin{eqnarray}
i\,\Pi_\Phi^{\mu\nu}(q^2,m^2,m^2)=\frac{ig^{\mu\nu}}{(4\pi)^2}\bigg[
2m^2(\Upsilon+1-\log\frac{m^2}{\mu^2})+\frac{q^2}{3}(-\Upsilon+\log\frac{m^2}{\mu^2})\bigg]
+(q^\mu q^\nu\mbox{~terms})+\mathcal{O}(q^4)\;.\label{iPi-Phi-2}
\end{eqnarray}
Eqs.\eqref{iPi-Omega}, \eqref{iPi-Phi} and \eqref{iPi-Phi-2} are
very practical in calculating the contributions of scalar loops to
vector boson self energies.

\begin{acknowledgments}
We would like to thank Q. Wang, J.~K.~Parry and R. Lu for insightful
discussions. This work is supported in part by the National Natural
Science Foundation of China.
\end{acknowledgments}


\begin{thebibliography}{99}
%\cite{Peskin:1990zt}
\bibitem{Peskin:1990zt}
  M.~E.~Peskin and T.~Takeuchi,
  %``A New constraint on a strongly interacting Higgs sector,''
  Phys.\ Rev.\ Lett.\  {\bf 65}, 964 (1990);
  %%CITATION = PRLTA,65,964;%%

%\cite{Peskin:1991sw}
%\bibitem{Peskin:1991sw}
  M.~E.~Peskin and T.~Takeuchi,
  %``Estimation of oblique electroweak corrections,''
  Phys.\ Rev.\ D {\bf 46}, 381 (1992).
  %%CITATION = PHRVA,D46,381;%%

%\cite{Yao:2006px}
\bibitem{Yao:2006px}
  W.~M.~Yao {\it et al.}  [Particle Data Group],
  %``Review of particle physics,''
  J.\ Phys.\ G {\bf 33}, 1 (2006).
  %%CITATION = JPHGB,G33,1;%%

%\cite{Georgi:1991ci}
\bibitem{Georgi:1991ci}
  H.~Georgi,
  %``Effective Field Theory And Electroweak Radiative Corrections,''
  Nucl.\ Phys.\  B {\bf 363}, 301 (1991).
  %%CITATION = NUPHA,B363,301;%%

%\cite{Dugan:1991ck}
\bibitem{Dugan:1991ck}
  M.~J.~Dugan and L.~Randall,
  %``The Sign Of S From Electroweak Radiative Corrections,''
  Phys.\ Lett.\ B {\bf 264}, 154 (1991).
  %%CITATION = PHLTA,B264,154;%%

%\cite{Gates:1991uu}
\bibitem{Gates:1991uu}
  E.~Gates and J.~Terning,
  %``Negative Contributions To S From Majorana Particles,''
  Phys.\ Rev.\ Lett.\  {\bf 67}, 1840 (1991).
  %%CITATION = PRLTA,67,1840;%%

%\cite{Luty:1992fe}
\bibitem{Luty:1992fe}
  M.~A.~Luty and R.~Sundrum,
  %``Technicolor theories with negative values of the Peskin-Takeuchi
  %electroweak parameter S,''
  Phys.\ Rev.\ Lett.\  {\bf 70}, 529 (1993)
  [arXiv:hep-ph/9209255].
  %%CITATION = PRLTA,70,529;%%


%\cite{Li:1992dt}
\bibitem{Li:1992dt}
  L.~F.~Li,
  %``Oblique electroweak corrections from heavy scalar fields,''
  Z.\ Phys.\ C {\bf 58}, 519 (1993).
  %%CITATION = ZEPYA,C58,519;%%

%\cite{Bhattacharyya:1993zy}
\bibitem{Bhattacharyya:1993zy}
  G.~Bhattacharyya, A.~Kundu, T.~De and B.~Dutta-Roy,
  %``Effects Of Isodoublet Color - Octet Scalar Bosons On Oblique Electroweak
  %Parameters,''
  J.\ Phys.\ G {\bf 21}, 153 (1995).
  %%CITATION = JPHGB,G21,153;%%

%\cite{Inami:1994nj}
\bibitem{Inami:1994nj}
  T.~Inami, T.~Kawakami and C.~S.~Lim,
  %``Constraints on the number of heavy generations from the S and T
  %parameters,''
  Mod.\ Phys.\ Lett.\ A {\bf 10}, 1471 (1995).
  %%CITATION = MPLAE,A10,1471;%%

%\cite{Bhattacharya:1994bj}
\bibitem{Bhattacharya:1994bj}
  G.~Bhattacharya, P.~Kalyniak and I.~Melo,
  %``Some constraints on neutral heavy leptons from flavor conserving decays of
  %the Z boson,''
  Phys.\ Rev.\ D {\bf 51}, 3569 (1995)
  [arXiv:hep-ph/9503248].
  %%CITATION = HEP-PH 9503248;%%

%\cite{Popovic:2000dx}
\bibitem{Popovic:2000dx}
  M.~B.~Popovic and E.~H.~Simmons,
  %``Weak-singlet fermions: Models and constraints,''
  Phys.\ Rev.\ D {\bf 62}, 035002 (2000)
  [arXiv:hep-ph/0001302].
  %%CITATION = HEP-PH 0001302;%%

%\cite{Peskin:2001rw}
\bibitem{Peskin:2001rw}
  M.~E.~Peskin and J.~D.~Wells,
  %``How can a heavy Higgs boson be consistent with the precision  electroweak
  %measurements?,''
  Phys.\ Rev.\  D {\bf 64}, 093003 (2001)
  [arXiv:hep-ph/0101342].
  %%CITATION = PHRVA,D64,093003;%%

%\cite{He:2001tp}
\bibitem{He:2001tp}
  H.~J.~He, N.~Polonsky and S.~f.~Su,
  %``Extra families, Higgs spectrum and oblique corrections,''
  Phys.\ Rev.\  D {\bf 64}, 053004 (2001)
  [arXiv:hep-ph/0102144].
  %%CITATION = PHRVA,D64,053004;%%


%\cite{Wells:2005vk}
\bibitem{Wells:2005vk}
  J.~D.~Wells,
  %``Introduction to precision electroweak analysis,''
  arXiv:hep-ph/0512342.
  %%CITATION = HEP-PH/0512342;%%


%\cite{Zhang:2006de}
\bibitem{Zhang:2006de}
  H.~H.~Zhang, Y.~Cao and Q.~Wang,
  %``The effects on S, T, and U from higher-dimensional fermion
  %representations,''
  arXiv:hep-ph/0610094.
  %%CITATION = HEP-PH 0610094;%%


\end{thebibliography}
\end{document}